# Reinforcement Learning from Demonstrations by Novel Interactive Expert and Application to Automatic Berthing Control Systems for Unmanned Surface Vessel

Haoran Zhang[a], Chenkun Yin[a]*, Yanxin Zhang[a], Shangtai Jin[a], Zhenxuan Li[b]

[a] *School of Electronic and Information Engineering, Beijing Jiaotong Univeristy, Shangyuancun 3#, Beijing, 100044, P.R.China*

[b] *Beijing Institute of Petrochemical Technology, Qingyuan North Road, Huangcun, Beijing, 102617, P.R.China*

**Abstract**

In this paper, two novel practical methods of Reinforcement Learning from Demonstration (RLfD) are developed and applied to automatic berthing control systems for Unmanned Surface Vessel. A new expert data generation method, called Model Predictive Based Expert (MPBE) which combines Model Predictive Control and Deep Deterministic Policy Gradient, is developed to provide high quality supervision data for RLfD algorithms. A straightforward RLfD method, model predictive Deep Deterministic Policy Gradient (MP-DDPG), is firstly introduced by replacing the RL agent with MPBE to directly interact with the environment. Then distribution mismatch problem is analyzed for MP-DDPG, and two techniques that alleviate distribution mismatch are proposed. Furthermore, another novel RLfD algorithm based on the MP-DDPG, called Self-Guided Actor-Critic (SGAC) is present, which can effectively leverage MPBE by continuously querying it to generate high quality expert data online. The distribution mismatch problem leading to unstable learning process is addressed by SGAC in a DAgger manner. In addition, theoretical analysis is given to prove that SGAC algorithm can converge with guaranteed monotonic improvement. Simulation results verify the effectiveness of MP-DDPG and SGAC to accomplish the ship berthing control task, and show advantages of SGAC comparing with other typical reinforcement learning algorithms and MP-DDPG.

## 1. Introduction

More and more complicated tasks for exploring in the oceans calls for strong demands of powerful unmanned surface vessels (USV), as well as reliable and intelligent automatic ship control systems [1]. Within all kinds of application scenarios, automatic berthing control for underactuated ships is considered to be one of the most important parts for an efficient and safe shipping task. However, restricted environment in the harbor and weak maneuverability of the ships at low speed bring great challenge to ship berthing, especially for underactuated ships. Therefore, it has significant importance to study and develop an efficient, accurate and reliable automatic berthing control system.

Classical automatic berthing system involves with sophisticated motion planning before conducing berthing task [2][3][4]. The control performance for berthing heavily relies on the planned route which is treated as the desired trajectory to be tracked by the controller [5][6][7]. Nevertheless, generating a safe geometric berthing path and corresponding time-dependent velocity (or position) trajectories is usually time-consuming and non-trivial. In addition, traditional berthing control method needs dynamic model of the ship, the modeling accuracy intrinsically influences the performance of the designed controller [8][9]. When some performance index to optimize is taken into consideration, such as minimum time or shortest path, the berthing task turns out to be a constrained nonlinear optimal control problem that need to solve Hamilton-Jacobi-Bellman equations [2][10].

* Corresponding author. Tel.:+86-13426277247. *E-mail address:* chkyin@bjtu.edu.cn.
This work is supported by National Natural Science Foundation of China under Grant 62073025 and 62073026.

Different from the numerical methods to solve this complicated HJB equations, Reinforcement Learning (RL) deals with the nonlinear optimal control problem under the framework of Markov Decision Process (MDP) [2][3][11]. One of the most important advantages of RL is the algorithm can work without any information about environment, hence control performance deteriorating caused by inaccurate dynamic modeling can be avoided. In addition, end-to-end RL enables the algorithm to learn the control policy directly without planning in advance. Hence, it is conceivable to apply RL methods to automatic ship berthing control task.

In recent years, RL attracts great attention in different fields with many successful applications including USVs [12]-[16][50]. In [14], two different RL algorithms, Deep Deterministic Policy Gradient (DDPG) and Normalized Advantage Function (NAF), are applied to the underactuated ship steering problem. However, only ship kinematics model is studied and only one control input, the turning rate, is considered in their work. [15] proposes a novel RL method based on policy iteration to track position trajectory for a fully actuated ship with known dynamics, and stability and convergence analysis is provided by Lyapunov method. In [16] static obstacle avoidance problem for USV is addressed via a combination of Deep Q-networks (DQN) [13] and specified avoidance reward function. A state-of-art RL method, Twin Delayed Deep Deterministic Policy Gradient (TD3), is successfully applied to motion planning for underactuated ship berthing task [50]. Simulation results show that TD3 outperform previous RL methods like DDPG in ship berthing task, especially when both kinematics and kinetics are taken into consideration.

However, unstable learning process and slow learning speed related with exploration problem of RL algorithm are the main problems from which typical RL algorithms always suffer. For example, simulations in the abovementioned papers are always conducted over thousands training *episodes* (iterations) before the algorithms converge [12][13][14][16]. To deal with the exploration problem, some researchers proposed Reinforcement Learning from Demonstration (RLfD) method, a combination of RL and Imitation Learning (IL), which utilizes the data collected by human experts or optimal controllers to assist the learning procedure. RLfD can significantly increases the sample efficiency with the guidance of the expert demonstrations to reduce unnecessary exploration. For IL, the expert policy is treated as the optimal one, under which the behaviour of experts from demonstrations is mimicked in a supervised learning manner. However, classical IL algorithms such as Data Aggregation (DAgger) [17] has no mechanism to optimize the agent policy according to the reward signals from interaction with the environment. Hence it is difficult for IL to drive the agent to output better performance than the expert.

Early RLfD methods have a relatively insufficient utilization of the expert data. Some of them utilize the property of off-policy RL method, such as DDPG and DQN, to directly load demonstrations into the replay buffer for improving the value estimation [20][21]. In [22], a model-free RL algorithm is initialized with a pre-trained policy by IL, and simulation results show that the learning speed can be accelerated. These methods that treat the demonstrations as augmentation of high-quality data cannot guarantee satisfactory mimicry of the expert policy especially when the expert data is deficient. In addition, distribution mismatch (drifting) is a typical shortcoming that limit the applications of RLfD, that is, the agent trajectories may drift away from expert trajectories because of compounding imitation error.

For efficient utilization of the expert demonstrations and alleviate distribution mismatch problem, many recent works on RLfD focus on integration of IL and RL. [23] introduces a RLfD method based on DAgger that learns the policy via minimizing the cost-to-go of an interactive expert, which is an expert that can be queried at any time during training. Instead of collecting expert trajectories in advance, DAgger-style methods attempt to deal with the distribution mismatch problem via continuously online querying the interactive expert to get expert actions of the state visited by the agent. However, it is difficult to apply due to lack of suitable interactive expert, especially human demonstrations. In [23] and [25], exploration is boosted via reward shaping by adding a supervision term to the typical RL objective function. In [25], generative adversarial network is utilized to describe the expert data distribution under the situation of few and sub-optimal demonstrations. In [29], an exploration region around the demonstrations is defined to reduce the side effect of sub-optimal expert demonstrations. Normalized Actor-Critic (NAC) proposed in [30] provides a unified approach for learning from the rewards and demonstrations while robustness to imperfect expert data is increased. However, the above RLfD methods all need human demonstrations.

There are also some algorithms utilize advanced controller to serve as the expert. Guided Policy Search (GPS) [26][27] and Policy Learning using Adaptive Trajectory Optimization (PLATO) [28], resort to optimal controllers such as model-predictive controller (MPC) [35] or LQG to generate supervision data. With help of these controllers, the agent shows good long-term performance with steadier and faster learning process than typical model-free RL algorithms. GPS handles distribution mismatch problem by modifying the objective function of the optimal controller such that the expert trajectories can be in the neighbourhood of agent trajectories, then optimizing the agent policy with respect to some supervised learning objective to match the expert policy. In the previous work [49], the authors proposed a novel DAgger-style RLfD method in which the demonstration data is generated online via a model-based RL method. In order to address the distribution mismatch problem, a heuristic mechanism was developed to balance the RL and IL according to a certain confidence measurement. However, the distribution mismatch problem is not sufficiently analysed in [49], and the heuristic updating law cannot be guaranteed to work stably and reliably. Compared to the previous work, the model-based RL method in [49] is refined here into an interactive expert which combines model-free RL and MPC mechanism more effectively. Then distribution mismatch problem that mentioned in [49] is theoretically analysed, and two techniques that can alleviate the problem are applied to making the algorithm more practical. In this work, we reformulate the RLfD problem as a dual optimization scheme to replace the heuristic mechanism that is less explainable and unstable. Furthermore, convergence analysis of the proposed RLfD method is given theoretically. The main contributions in this paper can be summarized as follows,

1) An interactive expert, Model Predictive Based Expert (MPBE), is proposed to generate high-quality demonstrations. An RLfD method called Model Predictive Deep Deterministic Policy Gradient (MP-DDPG), is extended from the model-based RL algorithm in [49] by utilizing the data collected from MPBE to update the agent.

2) The RL problem is reformulated as a constrained optimization problem to improve the learning performance of the agent via imitation from the expert demonstration given by MPBE. Another novel RLfD method called Self-Guided Actor-Critic (SGAC) is proposed to solve the constrained optimization problem in a DAgger-style and to avoid distribution mismatch problem. And theoretical analysis of SGAC is provided.

3) The UAV automatic berthing control task is accomplished by the proposed methods, and performance comparison is conducted between RLfD methods and typical model-free RL method TD3 [49].

The rest of the paper is organized as follows. The mathematical model of the underactuated ship, berthing control problem formulation and basic concepts of RL are introduced in section 2. In section 3, background of DDPG is firstly discribed, then MPBE is developed by combining DDPG with MPC. At last, distribution mismatch problem is theoretical analysed, which is followed by two modification technologies to enhance MP-DDPG algorithm. SGAC algorithm and convergence proof are detailed in section 4. In section 5, the ship berthing control problem is simulated by the proposed SGAC. The simulation results of SGAC and its comparison with Twin Delayed DDPG (TD3) and MP-DDPG are given before the conclusions.

## 2. Problem Formulation of Vessel Berthing Control and Preliminaries of RL

In this section, the dynamic model of an underactuated ship for berthing is described firstly. Then the automatic berthing task is formulated as a control problem under the framework of Markov Decision Process to be solved by RL methods. Necessary notations and preliminaries of RL is provided at the end of the section.

The kinematics model of a surface vessel is simply described by [1][8],

$$\dot{\eta} = J(\psi)\upsilon \tag{1}$$

where the vector $\eta = [X, Y, \psi]^T$ is consist of the ship's positions and yaw angle (orientation) in the earth-fixed reference frame, the velocity vector, $\upsilon = [u, v, \phi]^T$ is consist of surge velocity $u$, sway velocity $v$ and yaw velocity $\phi$ in the body-fixed reference frame, as depicted in Fig. 1.

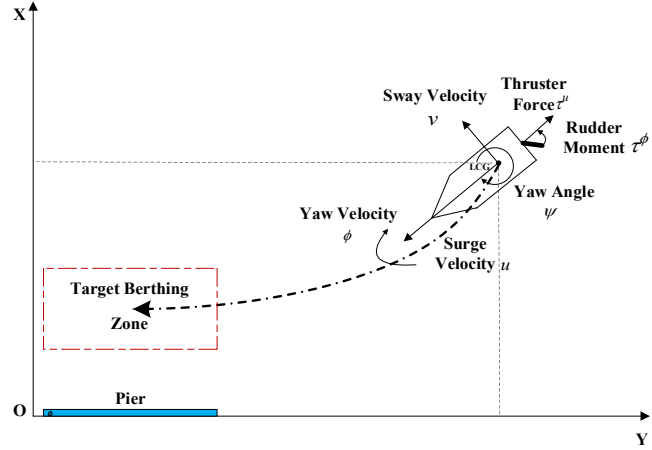

Figure 1. Ship dynamics for berthing

And, the kinematic transformation matrix $J(\psi)$ between the two frames is,

$$J(\psi) = \begin{bmatrix} \cos\psi & -\sin\psi & 0 \\ \sin\psi & \cos\psi & 0 \\ 0 & 0 & 1 \end{bmatrix}. \tag{2}$$

The following kinetics model is widely used for underactuated ships which has fewer control input variables than its degrees of freedom (DOF) [1][10],

$$M\dot{\upsilon} + C(\upsilon)\upsilon + D(\upsilon)\upsilon + g(\eta) = \tau + \tau_d. \tag{3}$$

Under the assumptions of low speed, ideal fluid and no incoming wave, the system inertia matrix $M$ can be treated as diagonal, non-singular and positive definite [8]. For simplicity, $M = \text{diag}\{m_{11}, m_{22}, m_{33}\}$ where $m_{11}$, $m_{22}$ and $m_{33}$ are positive constants. Coriolis and centripetal matrix $C(\upsilon)$ is skew-symmetric [9],

$$C(\upsilon) = \begin{bmatrix} 0 & 0 & -m_{22}v \\ 0 & 0 & m_{11}u \\ m_{22}v & -m_{11}u & 0 \end{bmatrix}. \tag{4}$$

The vector of restoring force $g(\eta)$ caused by gravity and buoyancy is ignored here since it is unnecessary to consider the heave, pitch, and roll dynamics of the ship affected by the restoring force for the berthing ship in the harbor [10].

For the control input vector $\tau = [\tau^u, 0, \tau^\phi]^T$ in the kinetics model (3), $\tau^u$ is the force in surge direction exerted by the thruster and $\tau^\phi$ is the moment in yaw direction exerted by the change of rudder angle. In this work, the environmental disturbance $\tau_d$ that reflects the impacts of wave, water flow and wind on the ship dynamics is neglected in still water environment.

By discretizing (1) and (3), the model of the underactuated surface vessel considered in this paper is summarized as below,

$$\begin{cases} u_{k+1} = u_k + \dfrac{m_{22}}{m_{11}} v_k \phi_k - \dfrac{d_{11}}{m_{11}} u_k + \dfrac{1}{m_{11}} \tau_k^u \\ v_{k+1} = v_k - \dfrac{m_{11}}{m_{22}} u_k \phi_k - \dfrac{d_{22}}{m_{22}} v_k \\ \phi_{k+1} = \phi_k + \dfrac{m_{11} - m_{22}}{m_{33}} u_k v_k - \dfrac{d_{33}}{m_{33}} \phi_k + \dfrac{1}{m_{33}} \tau_k^\phi \\ X_{k+1} = X_k + \cos(\psi_k) u_k - \sin(\psi_k) v_k \\ Y_{k+1} = Y_k + \sin(\psi_k) u_k + \cos(\psi_k) v_k \\ \psi_{k+1} = \psi_k + \phi_k. \end{cases} \quad (5)$$

In RL, the dynamic model is treated as the black-box environment that is an essential component. The agent has no access to the information about the dynamic model but only the data generated by interaction with the environment.

To accomplish the berthing task, the ship needs to gradually approach to the expected final position and moor in the berthing zone from the initial state. In this paper, the ship berthing problem is formulated as a discrete-time optimal control problem that maximizes a specified objective function subject to the ship dynamics (5) and other constraints.

Under the RL framework, the berthing control problem is considered as a standard MDP. For the MDP, a well-defined tuple ($S$, $A$, $P$, $R$, $\gamma$) is always needed. $S$ is a discrete-time continuous *state space* where the state $s_k = [u_k, v_k, \phi_k, X_k, Y_k, \psi_k]^T \in S$ is bounded by $s_{\min}$ and $s_{\max}$. $A$ is a discrete-time continuous *action space* in which the action $a_k = [\tau_k^u, \tau_k^\phi]^T \in A$ is bounded by $a_{\min}$ and $a_{\max}$, the limitation vectors for thruster and rudder angle. $P(s_{k+1}|s_k, a_k)$ is the *state-transition distribution* that describes the dynamics of a stochastic environment. The ship dynamics (5) or its compact form $s_{k+1} = f(s_k, a_k)$ can be treated as the *deterministic state-transition function*, which is a special case of $P(s_{k+1}|s_k, a_k)$ for a non-stochastic environment. $R(s_k, a_k): S \times A \to \mathbb{R}$ is the reward function which maps the state and agent's action to a scalar *reward* $r_k$. And $\gamma \in (0, 1]$ is a discount factor. In RL, the agent chooses an action $a_k$ according to a *policy* $\pi$ in a policy space $\Pi$, then receives a reward $r_k$ and next state $s_{k+1}$ from the environment.

In this paper, the agent to learn is a parameterized policy $\pi_\theta(a_k | s_k) \in \Pi$ which is differentiable w.r.t. the parameter $\theta$. The objective of RL is to find the optimal policy that maximizes the expected sum of discounted rewards as follows,

$$J(\pi_\theta) \triangleq J(\theta) = Q^{\pi_\theta}(s_0, a_0) = E_{a_{\kappa+1} \sim \pi_\theta, s_{\kappa+1} \sim P}[\sum_{\kappa=0}^{\infty} \gamma^\kappa R(s_\kappa, a_\kappa) | s_0, a_0], \quad (6)$$

where $Q^{\pi_\theta}(\cdot, \cdot)$ is the *action-value function* under a policy $\pi_\theta$, $P$ is $P(s_{k+1}|s_k, a_k)$ for short. Here the discount factor $\gamma$ guarantees the boundness of $J$ and $Q^{\pi_\theta}$.

The action-value $Q^{\pi_\theta}(s_k, a_k) = E_{a_{\kappa+1} \sim \pi_\theta, s_{\kappa+1} \sim P}[\sum_{\kappa=k}^{\infty} \gamma^{\kappa-k} R(s_\kappa, a_\kappa) | s_k, a_k]$ satisfies the self-consistency property that leads to the following Bellman equation [32][33][34],

$$Q^{\pi_\theta}(s_k, a_k) = r_k + \gamma E_{s_{k+1} \sim P}[Q^{\pi_\theta}(s_{k+1}, a_{k+1})]. \tag{7}$$

Another value function called *state-value* function $V^{\pi_\theta}(s_k)$ that evaluates performance of the policy $\pi_\theta(a_k | s_k)$ starting from the current state $s_k$ is defined as

$$V^{\pi_\theta}(s_k) = E_{a_\kappa \sim \pi_\theta, s_{\kappa+1} \sim P}[\sum_{\kappa=k}^{\infty} \gamma^{\kappa-k} R(s_\kappa, a_\kappa) | s_k]. \tag{8}$$

Subtracting the action-value by the state-value gives an advantage function $A^{\pi_\theta}(s_k, a_k)$ that evaluates the consequence by taking a certain action $a_k$ at current state $s_k$,

$$A^{\pi_\theta}(s_k, a_k) = Q^{\pi_\theta}(s_k, a_k) - V^{\pi_\theta}(s_k). \tag{9}$$

In order to facilitate of theoretical analysis, the discounted visitation frequency $\rho_{\pi_\theta}(s)$, which characterizes the unnormalized distribution of certain state within trajectories when $\pi_\theta$ is executed, is defined as follows [25][34],

$$\rho_{\pi_\theta}(s) = P_{\pi_\theta}(s_0 = s) + \gamma P_{\pi_\theta}(s_1 = s) + \gamma^2 P_{\pi_\theta}(s_2 = s) + ... = \sum_{k=0}^{\infty} \gamma^k P_{\pi_\theta}(s_k = s) \tag{10}$$

In the RL field, policy optimization procedure is usually implemented in an iterative style, it is reasonable to compare policy performance between consecutive iterations to analyze convergence of the algorithm. A useful conclusion [25][34], which connects any two policy parameters $\theta$ and $\theta'$ in terms of the performance $J(\cdot)$, is expressed as,

$$J(\theta') = J(\theta) + \int_s \rho_{\pi_{\theta'}}(s) \int_a \pi_{\theta'}(a|s) A^{\pi_\theta}(s,a). \tag{11}$$

Generally speaking, $\theta$ is the fixed policy parameter at current iteration while $\theta'$ is treated as the parameter to be updated by policy gradient method. However, the discounted visitation frequencies $\rho_{\pi_{\theta'}}(s)$ in (11) can neither be calculated when the environment dynamics $P$ is unknown, nor be well approximated by data during policy parameter updating. Therefore, a *surrogate objective* of $J(\cdot)$ is introduced instead [34],

$$\hat{J}_\theta(\theta') = J(\theta) + \int_s \rho_{\pi_\theta}(s) \int_a \pi_{\theta'}(a|s) A^{\pi_\theta}(s,a) \tag{12}$$

where the subscript $\theta$ denotes that $\hat{J}_\theta(\cdot)$ can be treated as the first order approximation of $J(\cdot)$ in the neighborhood of $\theta$ in the view of the following properties [34],

$$\begin{aligned} \hat{J}_\theta(\theta) &= J(\theta), \\ \nabla_{\theta'} \hat{J}_\theta(\theta') |_{\theta'=\theta} &= \nabla_{\theta'} J(\theta') |_{\theta'=\theta}. \end{aligned} \tag{13}$$

When policy gradient method is applied to update the policy parameter, it is convenient to calculate the gradient of the surrogate objective $\hat{J}_\theta(\cdot)$ via sample data as the following approximate gradient,

$$\nabla_\theta \hat{J}_\theta(\theta) = \nabla_\theta E_{(s_0, a_0, s_1, a_1, ...)}[Q^{\pi_\theta}(s_k, a_k)]. \tag{14}$$

where the samples $s_0, a_0, s_1, a_1, ....$ are collected by executing the policy $\pi_\theta$. Since the surrogate objective is a local approximation of the true objective $J(\cdot)$, one may optimize $\hat{J}_\theta(\cdot)$ via parameter updating $\theta \to \theta'$ with the approximate gradient (14) in replace of maximizing the objective function $J(\cdot)$ directly [11]. A useful result which reveals the relationship between $\hat{J}_\theta(\cdot)$ and $J(\cdot)$ is introduced in the following lemma.

**Lemma 1** [34]: Let $\varepsilon_\pi = \max_{s \in \mathbf{S}, a \in \mathbf{A}} |A^{\pi_\theta}(s,a)|$, the following bound holds for any two policy parameter $\theta$ and $\theta'$

$$J(\theta') \geq \hat{J}_\theta(\theta') - \frac{4\gamma\varepsilon_\pi}{(1-\gamma)^2}\max_s D_{KL}(\pi_{\theta'},\pi_\theta). \tag{15}$$

where $D_{KL}(\pi_{\theta'},\pi_\theta) = \int \pi_{\theta'}(a|s)\log(\pi_{\theta'}(a|s)/\pi_\theta(a|s))da$ is the Kullback–Leibler (KL) divergence between the distributions $\pi_{\theta'}$ and $\pi_\theta$.

## 3. Model Predictive Deep Deterministic Policy Gradient

For maximizing the RL objective (6) subject to (5) and other constraints, RLfD is an effective method that leverages expert demonstrations to accelerate the learning process. In this section, we propose a novel RLfD method by extending from DDPG, called MP-DDPG algorithm. A model predictive based expert, MPBE for shot, that provides demonstrations for MP-DDPG is firstly developed to aid the agent exploration. Then the distribution mismatch problem of MP-DDPG is theoretically analysed and two techniques are introduced to alleviate it.

### 3.1. Background of DDPG

Since DDPG is one of the most popular RL methods nowadays, only brief introduction is given here. And more comprehensive presentation of the Algorithm can be found in [33]. To find the optimal policy that maximizes $Q^{\pi_\theta}(s_k,a_k)$, DDPG takes advantage of *actor-critic* architecture and iteratively optimizes the agent policy by the gradient of an estimated action-value w.r.t. the policy parameter $\theta$ [33],

$$\mu_\theta(s_k) = \arg\max_\theta Q_\omega^{\pi_\theta}(s_k,a_k), \tag{16}$$

where $Q_\omega^{\pi_\theta}(s_k,a_k)$ is a critic-network parameterized by $\omega$ that estimates the action-value function $Q^{\pi_\theta}(s_k,a_k)$ via Deep Q-Network (DQN) [13], and $\mu_\theta(s_k)$ is an actor-network for approximating the optimal deterministic policy that can maximize $Q_\omega^{\pi_\theta}(s_k,a_k)$. In order to collect adequate data via interacting with the environment during training, the stochastic policy $\pi_\theta(a_k|s_k)$ is constructed as follows,

$$\pi_\theta(a_k|s_k) = \mu_\theta(s_k) + Z, \tag{17}$$

where $Z$ is sampled from Gaussian distribution $N(0,\Sigma_\pi)$. Therefore, $\pi_\theta$ is a Gaussian distribution with mean $\mu_\theta(s_k)$ and covariance matrix $\Sigma_\pi$. The deterministic policy $\mu_\theta(s_k)$ derived by (16) will be applied during testing and real-time applications after training.

In order to learn the parameters of the two neural networks, $\theta$ and $\omega$ respectively, DDPG randomly samples a batch of historical experiences $(s_i,a_i,r_i,s_{i+1})$ from a replay buffer (RB) and updates the actor-network in the following manner,

$$\theta' = \theta + \alpha_{DDPG} \cdot E_{s_i \sim RB}[\nabla_\mu Q_\omega^{\pi_\theta}(s_i,\mu_\theta(s_i)) \cdot \nabla_\theta \mu_\theta(s_i)] \tag{18}$$

where $\theta'$ is the actor-network parameter updated in the next learning episode, and the expectation in the R.H.S of (18) is an approximation of the true gradient $\nabla_\theta J(\theta)$ using data stored in RB as mentioned in last section. By utilizing the same experiences from RB, the critic-network is trained by minimizing the *temporal-difference* (TD) error between the TD-target $y_i$ and the current estimation of $Q_\omega^{\pi_\theta}(s_i,a_i)$ as follows,

$$\omega = \arg\min_\omega E_{(s_i,a_i,r_i,s_{i+1}) \sim RB}[(y_i - Q_\omega^{\pi_\theta}(s_i,a_i))^2], \quad (19)$$
$$y_i = r_i + \gamma Q_{\bar{\omega}}^{\pi_\theta}(s_{i+1}, \mu_{\bar{\theta}}(s_{i+1})).$$

where $\bar{\theta}$ and $\bar{\omega}$ denote parameters of two separate *target networks*. By following the idea of DQN, the target networks are copies of the actor-network and the critic-network respectively to generate the TD-target value in (19). Meanwhile, updating of these weights in the target networks is delayed as follows in order to make the learning process of the critic-network stable,

$$\bar{\theta} \leftarrow \varepsilon\theta + (1-\varepsilon)\bar{\theta}, \quad \bar{\omega} \leftarrow \varepsilon\omega + (1-\varepsilon)\bar{\omega}, \quad (20)$$

where $\varepsilon \ll 1$.

Although DDPG can learn optimal policy for complex control task, it still suffers from the exploration problem. The agent with the policy (17), however, is incapable of exploring the environment entirely, thus the learned policy may be vulnerable to disturbances and unfamiliar states. Another shortcoming of DDPG is that the learning process cannot be guaranteed to converge. The problem cause by weak exploration usually leads to serious performance deterioration during the learning process especially when the critic-network fails to estimate $Q_\omega^{\pi_\theta}(s_i,a_i)$ accurately.

*3.2. Algorithm Design of MP-DDPG with the aid of MPBE*

MPC is considered as one of the most powerful methods to solve complex control problems [35]. A common nonlinear MPC is consist of a predictive model, a controller and an optimizer. At each time step, the controller interacts with the predictive model to outcome control sequences within finite time horizon, while the optimizer calculates optimal control actions in a receding horizon manner. The very first action from the optimal control sequence will be implemented as the control input, then MPC is recurrently carried out until some stopping criteria is satisfied [36]. And some recent works discovered underlying relationships between RL and MPC [37]-[40]. In order to improve the sample efficiency and robustness of RL, we combine RL with the prediction and optimization mechanism of MPC.

In this section, a novel RLfD algorithm, namely MP-DDPG, is developed to accelerate the learning speed of the prototype DDPG. Firstly, an Expert Demonstration Data Generating (MPBE) method is investigated. In MPBE, a neural network based environment model, $\hat{f}(s_k,a_k)$, serves as a simulator to be interacted with the agent at every time step $k$ to generate $M$ different control sequences $U_k^M = (A_{k:k+H}^{(1)},\ldots,A_{k:k+H}^{(M)})$ over a finite optimization horizon $H$. $A_{k:k+H}^{(m)} = (a_k^{(m)},\ldots,a_{k+H}^{(m)})$ denotes the predictive control sequence which contains $H$ actions, which are sampled from $\pi_\theta$ as follows,

$$a_l^{(m)} \sim \pi_\theta = \mu_\theta(\hat{s}_l^{(m)}) + Z, \quad m=1,\ldots,M, \quad l=k,\ldots,k+H, \quad (21)$$

where the state and its predictions are

$$\hat{s}_k^{(m)} = s_k,$$
$$\hat{s}_{l+1}^{(m)} = \hat{s}_l^{(m)} + \hat{f}(\hat{s}_l^{(m)}, a_l^{(m)}), \quad l=k+1,\ldots,k+H. \quad (22)$$

$s_k$ is the current state received from the environment, and the *rollout* $\tau_k^{(m)}: s_k, a_k^{(m)},\ldots,\hat{s}_{k+H-1}^{(m)}, a_{k+H-1}^{(m)}, \hat{s}_{k+H}^{(m)}$ is predicted by $\hat{f}(s_k,a_k)$. Here, the neural network environment model $\hat{f}$ is iteratively trained along with the actor and critic networks of the agent from the data in the replay buffer: $\hat{f}(\cdot,\cdot) = \min_{\hat{f} \in F} E_{(s_i,a_i,s_{i+1}) \sim RB}[\|s_{i+1} - \hat{f}(s_i,a_i)\|_2]$. If the system parameters are known, the mathematical model (5) can be directly used as $\hat{f}$ here. Unlike model-free RL methods, the additional model $\hat{f}$ can be used to accelerate policy learning in a receding horizon optimization manner as described below.

Then an elite control sequence $A_{k:k+H}^E$ can be selected by sorting $U_k^M$ as follows,

$$A^E_{k:k+H} = \underset{A^{(m)}_{k:k+H} \in U^M_k}{\arg\max} \; \widehat{G}(s_k, a_k^{(m)}) \triangleq \underset{A^{(m)}_{k:k+H} \in U^M_k}{\arg\max} \left( \sum_{l=k}^{k+H-1} \gamma^{l-k} R(\hat{s}_l^{(m)}, a_l^{(m)}) + \gamma^H Q^{\pi_\theta}_{\bar{\omega}}(\hat{s}_H^{(m)}, a_H^{(m)}) \right), \quad (23)$$

where the sum of discounted rewards from $\hat{s}_H^{(m)}, a_H^{(m)}$ till the end of a rollout is estimated by $Q^{\pi_\theta}_{\bar{\omega}}(\hat{s}_H^{(m)}, a_H^{(m)})$.

Instead of executing the whole control sequence $A^E_{k:k+H}$, only the first action $a^E_k$ in $A^E_{k:k+H}$ is applied to the environment. This is similar with random shooting method [36], a typical technique in nonlinear MPC. However, $a_l^{(m)}$ is generated by current policy $\pi_\theta$ instead of random sampling in the action space $A$. Since each $a_l^{(m)}$ is sampled from a Gaussian policy $\pi_\theta$ as in (21), it is reasonable to assume that $a^E_k$ satisfies another Gaussian distribution $\pi_E(a^E_k | s_k)$ with an unknown mean $\mu_E(s_k)$ and a covariance matrix $\Sigma_E$.

The MPBE procedures are summarised in Algorithm 1. The loop from step 3-7 can generate a rollout $\tau_k^{(m)} : s_k, a_k^{(m)}, \ldots, \hat{s}_{k+H-1}^{(m)}, a_{k+H-1}^{(m)}, \hat{s}_{k+H}^{(m)}$ by interaction between the agent and the model $\hat{f}$. In addition, the return $\widehat{G}(s_k, a_k^{(m)})$ of this rollout is calculated according to the reward function and the target critic-network $Q^{\pi_\theta}_{\bar{\omega}}$ in step 8. In step 11, the elite control sequence $A^E_{k:k+H}$ is obtained by selecting the sequence with highest return $\widehat{G}(s_k, a_k^{(m)})$. In the final step, the first element of $A^E_{k:k+H}$ is treated as the expert action $a^E_k$.

**Lemma 2** [41]: Given two $n$ dimensional Gaussian policy distributions $\pi_E$ and $\pi_\theta$ with means $\mu_\theta, \mu_E$ and covariance matrices $\Sigma_\pi, \Sigma_E$ respectively, we have,

$$D_{KL}(\pi_E, \pi_\theta) = \frac{1}{2}(\log \frac{|\Sigma_\pi|}{|\Sigma_E|} - n + tr(\Sigma_\pi^{-1} \Sigma_E) + (\mu_\theta - \mu_E)^T \Sigma_\pi^{-1} (\mu_\theta - \mu_E)). \quad (24)$$

Now it is ready to develop MP-DDPG by incorporating MPBE with DDPG. As shown in Figure 2, MP-DDPG makes full use of the expert action $a^E_k$ from MPBE to generate high-quality training transitions $(s^E_k, a^E_k, r^E_k, s^E_{k+1})$ by interacting with the environment directly. Then the transition $(s^E_k, a^E_k, r^E_k, s^E_{k+1})$ is uploaded to replay buffer (RB) at each time-step. Therefore $(s^E_k, a^E_k, r^E_k, s^E_{k+1})$ can be treated as the expert demonstration at time $k$ for updating the networks using the same techniques as DDPG. In this point of view, MP-DDPG has common characteristic similar with Deep Deterministic Policy Gradient from Demonstrations (DDPGfD) [20]. By learning from the expert demonstration $(s^E_k, a^E_k, r^E_k, s^E_{k+1})$, the convergence speed of the critic-network would be accelerated.

MP-DDPG directly leverages expert data $(s^E_k, a^E_k, r^E_k, s^E_{k+1})$ stored in the replay buffer to update the critic-network by minimizing the following lost function,

$$L_{fD}(\omega) = E_{(s^E_i, a^E_i, r^E_i, s^E_{i+1}) \sim RB}[(r^E_i + \gamma Q_{\bar{\omega}}(s^E_{i+1}, \mu_{\bar{\theta}}(s^E_{i+1})) - Q_\omega(s^E_i, a^E_i))^2]. \quad (25)$$

**Assumption 1**: The expert policy $\pi_E$ has a non-negative advantage value over the current agent policy $\pi_\theta$,

$$E_{a_k \sim \pi_E}[A^{\pi_\theta}(s_k, a_k)] = E_{a_k \sim \pi_E}[Q^{\pi_\theta}(s_k, a_k) - V^{\pi_\theta}(s_k)] = \sigma, \forall s_k \in S, \quad (26)$$

where $\sigma \geq 0$.

Assumption 1 indicates that although the expert policy $\pi_E$ may not be the global optimal policy, but it is better than the agent policy with the current parameter $\theta$. This assumption is reasonable since $a_k^E$ is calculated after the optimization procedure in (23), and $Q^{\pi_\theta}(s_k, a_k^E)$ should not be smaller than the average one $V^{\pi_\theta}(s_k)$ at the same state $s_k$.

**Remark 1**: The difference between MP-DDPG and DDPGfD is significant, the latter one requires collecting expert data from a human demonstrator in advance, while MP-DDPG employs an adaptive expert $\pi_E$ to generate demonstrations online. Once the actor-network is updated, all sequences in $U_k^M$ may have better performance. It means that the expert from MP-DDPG can adaptively improve its own performance during the training process.

---

**Algorithm 1:** Model Predictive based Expert (MPBE)

---

**Initialize:** neural network dynamic model $\hat{f}$, optimization horizon $H$ and the number of control sequences $M$

**Call:** actor-network $\mu_\theta$, target critic-network $Q_\omega^{\pi_\theta}$ and experience replay buffer $RB$

1: **Input:** current state $s_k$
2: **for** control sequence $m = 1, ..., M$ **do:**
3:     **for** prediction step $l = k, k+1, ..., k+H$ **do:**
4:         **Compute** action $a_l^{(m)} \sim \pi_\theta = \mu_\theta(\hat{s}_l^{(m)}) + Z$ by (21)
5:         **Predict** next state by dynamic model $\hat{s}_{l+1}^{(m)} = \hat{s}_l^{(m)} + \hat{f}(\hat{s}_l^{(m)}, a_l^{(m)})$ by (22)
6:         **Step forward** $l = l+1$
7:     **end for**
8:     **Obtain** the $m$-th control sequence $A_{k:k+H}^{(m)} = (a_k^{(m)}, ..., a_{k+H}^{(m)})$ and return $\hat{G}(s_k, a_k^{(m)})$
9: **end for**
10: **Obtain** $M$ different control sequences $U_k^M = (A_{k:k+H}^{(1)}, ..., A_{k:k+H}^{(M)})$
11: **Compute** an elite control sequence $A_{k:k+H}^E$ by (23)
12: **Output:** expert action $a_k^E$ from $A_{k:k+H}^E$

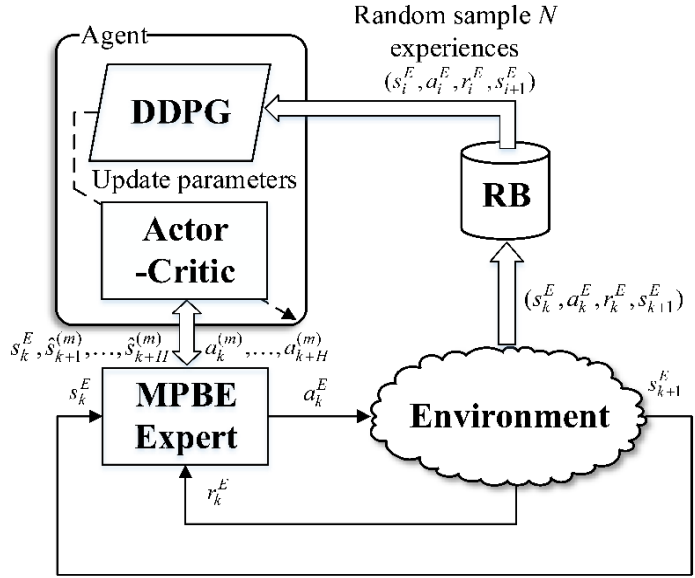

Figure 2. Block Diagram of MP-DDPG

*3.3. Distribution Mismatch Problem of MP-DDPG*

In this section, the distribution mismatch problem of MP-DDPG is analysed, then two modification techniques are applied to MP-DDPG. We first define the following action-value Bellman operator $\mathcal{T}^{\pi_E}$ used in MP-DDPG according to (25)

$$\mathcal{T}^{\pi_E} Q(s,a) \triangleq R(s,a) + \gamma E_{s' \sim P(s'|s,\pi_E), a' \sim \pi_\theta}[Q(s',a')], \forall s \in \mathbf{S}, \forall a \in \mathbf{A}. \tag{27}$$

where $s', a'$ are the subsequent state and action of $s, a$ respectively. Next, we prove that the critic-network in MP-DDPG can estimate the action-value $Q^{\pi_\theta}$ through the operator $\mathcal{T}^{\pi_E}$ if certain constraint on the agent policy and the expert policy is satisfied. Otherwise, the distribution mismatch problem leading to poor performance or divergence of the learned policy is inevitable.

**Theorem 1**: For given gaussian policies $\pi_\theta$ and $\pi_E$, the operator $\mathcal{T}^{\pi_E}$ used in MP-DDPG and a mapping $Q_e : \mathbf{S} \times \mathbf{A} \to \mathbb{R}, e = 0, 1, ...,$ the sequence of $Q_{e+1} \triangleq \mathcal{T}^{\pi_E} Q_e$ will converge to the true action-value $Q^{\pi_\theta}$ of the agent policy,

$$\mathcal{T}^{\pi_E} .... \mathcal{T}^{\pi_E} \mathcal{T}^{\pi_E} Q_0 \to Q^{\pi_\theta}, \tag{28}$$

if KL divergence of trajectory distributions between the expert and the agent is zero, namely

$$D_{KL}(p(\tau_E), p(\tau_\theta)) = 0. \tag{29}$$

**Proof:** According to the definition of KL divergence, we have

$$D_{KL}(p(\tau_E), p(\tau_\theta))$$
$$= E_{p(\tau_E)}[\log p(\tau_E) - \log p(\tau_\theta)]$$
$$= E_{p(\tau_E)}\left[\log \frac{p(s_0)\prod_{k=0}^{\infty} \pi_E(a_k^E | s_k)P(s_{k+1} | s_k, a_k^E)}{p(s_0)\prod_{k=0}^{\infty} \pi_\theta(a_k | s_k)P(s_{k+1} | s_k, a_k)}\right]$$
$$= \sum_{k=0}^{\infty} E_{\pi_E}\left[\log \frac{\pi_E(a_k^E | s_k)}{\pi_\theta(a_k | s_k)}\right]$$
$$= \sum_{k=0}^{\infty} D_{KL}(\pi_E(a_k | s_k), \pi_\theta(a_k | s_k)).$$

Based on this result and the property of KL divergence [41], the following conclusions are straightforward under the condition $D_{KL}(p(\tau_E), p(\tau_\theta)) = 0$,

$$p(\tau_E) = p(\tau_\theta),$$
$$\pi_E(a^E | s) = \pi_\theta(a | s), \forall s \in \mathbf{S}.$$

Then

$$\mathcal{T}^{\pi_E} Q^{\pi_\theta}(s,a) = R(s,a) + \gamma E_{s' \sim P(s'|s,\pi_E), a' \sim \pi_\theta}[Q^{\pi_\theta}(s',a')]$$
$$= R(s,a) + \gamma \int_{s'} P(s'|s,\pi_E) \int_{a'} \pi_\theta(a'|s')Q^{\pi_\theta}(s',a')$$
$$= R(s,a) + \gamma \int_{s'} P(s'|s,\pi_\theta) \int_{a'} \pi_\theta(a'|s')Q^{\pi_\theta}(s',a')$$
$$= Q^{\pi_\theta}(s,a)$$

This means $Q^{\pi_\theta}$ is a fixed point of the operator $\mathcal{T}^{\pi_E}$.

Next, we prove that $\mathcal{T}^{\pi_E}$ is a $\gamma$-contraction. For any $Q_a$ and $Q_b$ from the sequence $\{Q_e\}_{e=0}$, we have,

$$\left\|\mathcal{T}^{\pi_E} Q_a - \mathcal{T}^{\pi_E} Q_b\right\|_\infty$$
$$= \left\|\gamma \int_{s'} P(s'|s,\pi_E) \int_{a'} \pi_\theta(a'|s')(Q_a(s',a') - Q_b(s',a'))\right\|_\infty$$
$$\leq \left\|\gamma \int_{s'} P(s'|s,\pi_E) \int_{a'} \pi_\theta(a'|s') \|Q_a(s',a') - Q_b(s',a')\|_\infty\right\|_\infty$$
$$\leq \gamma \|Q_a(s',a') - Q_b(s',a')\|_\infty, \gamma \in [0,1).$$

Hence, $\mathcal{T}^{\pi_E}$ is a $\gamma$-contraction.

At last, we show the sequence $\{Q_e\}_{e=0}$ converges to $Q^{\pi_\theta}$ as $e \to \infty$,

$$\left\|Q_{e+1} - Q^{\pi_\theta}\right\|_\infty = \left\|\mathcal{T}^{\pi_E} Q_e - \mathcal{T}^{\pi_E} Q^{\pi_\theta}\right\|_\infty \leq \gamma \left\|Q_e - Q^{\pi_\theta}\right\|_\infty \leq ... \leq \gamma^{e+1} \left\|Q_0 - Q^{\pi_\theta}\right\|_\infty \to 0. \square$$

In prototype MP-DDPG as introduced aforementioned and shown in Fig.2, no mechanism is applied to guarantee the strict condition (29) in Theorem 1. It means the critic-network of MP-DDPG inevitably generates biased estimation of $Q^{\pi_\theta}$. As a result, MP-DDPG has to suffer from distribution mismatch problem

(drifting) which may cause poor performance or divergence of the learned policy from the expert one. Formally speaking, the distribution mismatch problem occurs when the distribution $p(\tau_E)$ of the expert trajectory $\tau_E$ is not the same as the distribution $p(\tau_\theta)$ of the agent trajectory $\tau_\theta$, namely that,

$$p(\tau_E) \neq p(\tau_\theta),$$

where

$$p(\tau_\theta) = p(s_0)\prod_{k=0}^{\infty} \pi_\theta(a_k | s_k)P(s_{k+1} | s_k, a_k),$$

$$p(\tau_E) = p(s_0)\prod_{k=0}^{\infty} \pi_E(a_k^E | s_k)P(s_{k+1} | s_k, a_k^E).$$

In order to alleviate the biased estimation problem of MP-DDPG, *stochastic mixing* (SM) technique similar with the one used in [42] can be conducted by alternatively executing $\pi_\theta$ and $\pi_E$ at each time step $k$ during training. And this leads to a new loss function of the critic-network of MP-DDPG,

$$L_{MP-DDPG}(\omega) = E_{(s_i^E, a_i^E, r_i^E, s_{i+1}^E) \sim RB_E}[(y_i^E - Q_\omega(s_i^E, a_i^E))^2] + E_{(s_i, a_i, r_i, s_{i+1}) \sim RB_\theta}[(y_i - Q_\omega(s_i, a_i))^2], \quad (30)$$

where $y_i^E = r_i^E + \gamma Q_{\bar{\omega}}(s_{i+1}^E, \mu_{\bar{\theta}}(s_{i+1}^E))$ and $y_i = r_i + \gamma Q_{\bar{\omega}}(s_{i+1}, \mu_{\bar{\theta}}(s_{i+1}))$. The data $(s_k, a_k, r_k, s_{k+1})$ generated by the agent policy $\pi_\theta$ is stored in on buffer $RB_\theta$, while $(s_k^E, a_k^E, r_k^E, s_{k+1}^E)$ generated by elite policy $\pi_E$ is stored in another buffer $RB_E$. This modification makes a trade-off between standard DDPG and MP-DDPG. If all data are generated by $\pi_\theta$, then the distribution mismatch problem can be eliminated and the algorithm becomes standard DDPG.

Moreover, a *behavioral cloning* (BC) technique [42] can also be applied. Specifically, a supervision term that encourages the agent to imitate expert behavior is added to the original objective function of actor-network (18) to deal with the distribution mismatch problem,

$$\hat{J}_{MP-DDPG}(\theta) = E_{s_i \sim RB_\theta}[Q_\omega(s_i, \mu_\theta(s_i))] - \lambda_{BC} \cdot E_{(s_i^E, a_i^E) \sim RB_E}[\| \mu_\theta(s_i^E) - a_i^E \|_2]. \quad (31)$$

**Remark 2:** BC technique used in (31) can be also seen as *policy distillation technology* to extract the knowledge from MP-DDPG into the actor-network [44]. A recent work that makes use of MPC and policy distillation is POPLIN [45]. However, POPLIN is a pure model-based method rather than any RLfD method as MP-DDPG.

Intuitively, BC is a supervised learning method that greedily maximizes the probability of the agent taking the expert action $a_i^E$ at an expert state $s_i^E$ [18]. However, violation of the important *i.i.d.* condition for the data generated by MDP may cause the collapse of supervised learning algorithms. More seriously, imitation error due to distribution mismatch is accumulating over time. A small mistake made by the agent can make the agent trajectory gradually drift to some unfamiliar states, and the agent has no ability to return back to the expert trajectory since the expert data cannot cover all possible states visited by the agent. In [17], performance divergence of policies between the agent and the expert caused by BC term is studied.

In summary, the pseudo code of MP-DDPG with SM & BC is given in Algorithm 2. The if-else branch from step 4 to 13 realizes the stochastic mixing (SM) technique by alternatively executing $\pi_\theta$ and $\pi_E$. In step 5, the expert action $a_k^E$ is computed by inputting the current state to MPBE as shown in the step 1 of Algorithm 1. In step 14, $N$ transitions of $(s_i^E, a_i^E, r_i^E, s_{i+1}^E)$ and $N$ $(s_i, a_i, r_i, s_{i+1})$ are random sampled from $RB_E$ and $RB_\theta$ respectively in order to compute loss function of the critic-network $L_{MP-DDPG}(\omega)$ as in (30). In step 15, MP-DDPG updates the critic-network and the actor-network with new objectives $L_{MP-DDPG}(\omega)$ (30) and $\hat{J}_{MP-DDPG}(\theta)$ (31) respectively via the same manner in DDPG (18), (19). In step 16, target networks are updated by (20). In step 18, the dynamic model is trained via samples from $RB_E$ and $RB_\theta$ as described in Section 3.2.

## 4. SGAC: A Modification of MP-DDPG by Utilization of Expert Data

*4.1. Algorithm Design*

Although MP-DDPG given in Section 3.3 utilizes several techniques like SM and BS to improve the learning performance of the agent policy, the distribution mismatch problem still exists since the condition (29) in Theorem 1 cannot be guaranteed by Algorithm 2. Hence, in order to address this problem thoroughly, the requirement on KL divergence $D_{KL}(\pi_E(a_k|s_k),\pi_\theta(a_k|s_k))=0$ at each time step $k$ is imposed directly for solving the policy optimization problem via maximizing (6) as,

$$\max_\theta J(\theta) = \max_\theta Q^{\pi_\theta}(s_0, a_0)$$
$$s.t.\ D_{KL}(\pi_E(a_k|s_k),\pi_\theta(a_k|s_k))=0, \forall k, \tag{32}$$

where $\pi_E(a_k|s_k)$ is provided by MPBE as used in MP-DDPG. For iteratively solving problem (32), it requires that the parameterized policy maximizes the original RL objective (6) while it needs to converge to the expert policy finally. By introducing the KL divergence constraint in (32), we not only address the

---

**Algorithm 2:** Model Predictive DDPG with SM & BC

**Initialize** critic-network $Q_\omega$ and actor-network $\mu_\theta$, target networks $Q_{\bar\omega}$ and $\mu_{\bar\theta}$ with $\bar\theta \leftarrow \theta$ and $\bar\omega \leftarrow \omega$, experience replay buffer $RB$ and $RB_\theta$, dynamic model $\hat{f}$, optimization horizon $H$ and the number of control sequences $M$, total time horizon $T$.

1: **for** each episode **do:**
2:    **for** time step $k = 0, \ldots, T$ **do:**
3:       **Observe** current state $s_k$
4:       **if** $k$ is odd **do:**
5:          **Obtain** expert action $a_k^E$ by MPBE
6:          **Execute** $a_k^E$ to the environment by (21)
7:          **Receive** reward $r_k^E$ and next state $s_{k+1}^E$ from the environment
8:          **Store** transition $(s_k^E, a_k^E, r_k^E, s_{k+1}^E)$ in $RB_E$
9:       **else:**
10:      **Execute** the agent action $a_k$
11:      **Receive** reward $r_k$ and next state $s_{k+1}$ from the environment
12:      **Store** transition $(s_k, a_k, r_k, s_{k+1})$ in $RB_\theta$
13:      **end if**
14:      **Sample** $N$ transitions $(s_i^E, a_i^E, r_i^E, s_{i+1}^E)$ and $(s_i, a_i, r_i, s_{i+1})$ from $RB_E$ and $RB_\theta$
15:      **Update** the critic-network $Q_\omega$ by minimizing (30)
16:      **Update** the actor-network $\mu_\theta$ by maximizing (31)
17:      **Update** the target networks by (20)
18:    **end for**
19:    **Update** the dynamic model $\hat{f}$
20: **end for** and **output** the trained policy $\pi_\theta$

distribution mismatch problem, but also reduce the random search at the early stage of training by forcing the agent to mimic the expert policy. Nevertheless, we still need to be clarified the following issues before developing a practical algorithm: 1) Analytical expression of the expert policy is unknown to the agent, which means the KL divergence can only be estimated by interaction data. 2) The expert data should cover all possible situation that the agent may encounter to prevent distribution mismatch problem. 3) The optimization mechanism needs to be designed carefully under the framework of both reinforcement learning and imitation learning for updating the agent policy.

In order to solve the constrained optimization problem (32) and to utilize expert experience efficiently, a novel RLfD algorithm called Self-Guided Actor-Critic (SGAC) is proposed here by extending from MP-DDPG. Different from updating the parameters passively by using expert data as collected in MP-DDPG, SGAC can actively query the MPBE to provide demonstration action for each state explored by the agent itself. Then the sampled data along with the expert action can be used to estimate the KL divergence and then the RLfD problem (32) is solved via *dual gradient optimization* [46] Although MPBE developed in Section 3.2 is utilized in both SGAC and MP-DDPG, it is served as an interactive expert here as shown in the Figure 3.

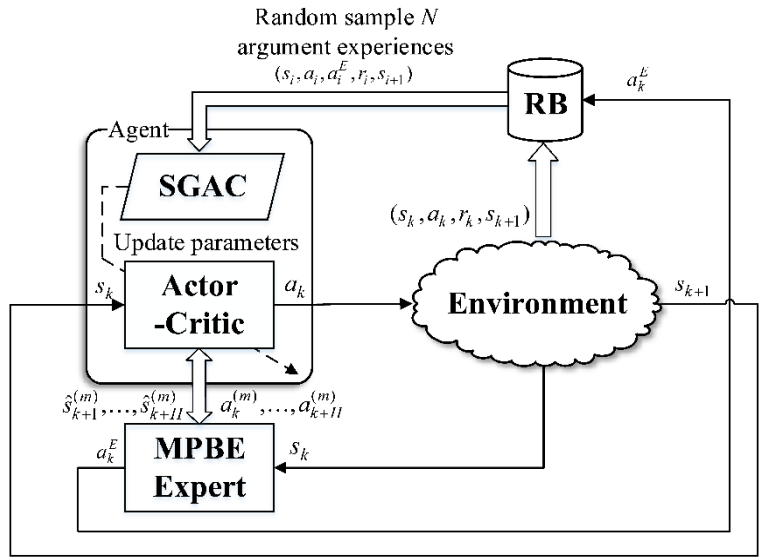

Figure 3. Block Diagram of SGAC

In SGAC, the problem (32) is rewritten as a minmax optimization problem,

$$\min_{\lambda} \max_{\theta} \eta(\theta,\lambda) = \min_{\lambda} \max_{\theta} [Q_{\omega}^{\pi_\theta}(s_0,a_0) - E_{s_k \sim p(\tau_\theta)}(\lambda \cdot D_{KL}(\pi_E(a_k | s_k), \pi_\theta(a_k | s_k)))], \quad (33)$$

where $\lambda \geq 0$ is the dual variable. SGAC algorithm for solving (32) is implemented by alternatively optimizing $\theta$ and updating dual variable $\lambda$.

On the one hand, we maximize a surrogate objective $\hat{\eta}$, i.e. the estimation of $\eta$, w.r.t $\theta$ by applying the gradient ascent algorithm as follows,

$$\theta' = \theta + \alpha_{SGAC} \cdot \nabla_\theta \hat{\eta}(\theta,\lambda), \quad (34)$$

where, $\alpha_{SGAC}$ is a step parameter, and

$$\begin{aligned}
\nabla_\theta \hat{\eta}(\theta, \lambda) &= \nabla_\theta E_{(s_i, a_i^E) \sim RB}[Q_\omega^{\pi_\theta}(s_i, \pi_\theta(s_i)) - \lambda \cdot D_{KL}(\pi_E, \pi_\theta)] \\
&= E_{(s_i, a_i^E) \sim RB}[\nabla_\theta Q_\omega^{\pi_\theta}(s_i, \pi_\theta(s_i)) - \lambda \cdot \nabla_\theta D_{KL}(\pi_E, \pi_\theta)] \\
&= E_{(s_i, a_i^E) \sim RB}[\nabla_\theta Q_\omega^{\pi_\theta}(s_i, \pi_\theta(s_i)) - \frac{1}{2}\lambda \cdot \nabla_\theta (\mu_\theta(s_i) - a_i^E)^T \Sigma_\pi^{-1}(\mu_\theta(s_i) - a_i^E)] \\
&= E_{(s_i, a_i^E) \sim RB, Z \sim N}[\nabla_\mu Q_\omega^{\pi_\theta}(s_i, \mu_\theta(s_i) + Z)\nabla_\theta \mu_\theta(s_i) - \frac{1}{2}\lambda \cdot \nabla_\theta (\mu_\theta(s_i) - a_i^E)^T \Sigma_\pi^{-1}(\mu_\theta(s_i) - a_i^E)]
\end{aligned} \quad (35)$$

where the expectation is computed over N random sampling of $(s_k, a_k^E)$ from *RB*, and the dual variable $\lambda$ is treated as a constant during optimizing $\theta$. The third equation of (35) follows from Lemma 1 since the terms independent of $\theta$ vanish by taking gradient. The last equation holds since the policy $\pi_\theta$ is reparametrized as (17). On the other hand, the following equation is applied for minimizing operation w.r.t $\lambda$ where $\theta'$ is treated as a constant obtained from (34),

$$\lambda' = \lambda - \zeta \cdot \nabla_\lambda \hat{\eta}(\theta', \lambda), \quad (36)$$

where,

$$\nabla_\lambda \hat{\eta}(\theta', \lambda) = -\frac{1}{2} E_{(s_i, a_i^E) \sim RB}[\| \mu_{\theta'}(s_i) - a_i^E \|_2]. \quad (37)$$

Comparing with the gradient (18) used in DDPG, the additional expert imitation term in (35) can accelerate the learning process as shown in [25]. In Figure 4, the elliptic region represents the policy space. Deeper color of the region means, the performance of the policy is better in terms of higher action value *Q*. The updating procedure of DDPG is illustrated by the purple arrows on the left side, while SGAC updating by the black arrows on the other side. With the help of the KL divergence term in SGAC, the agent converges to the optimal policy with fewer iterations than DDPG.

**Remark 3:** Although SGAC and MP-DDPG are both RLfD methods, one key difference between them lies in, all states of SGAC are visited by the agent policy $\pi_\theta$, rather than the elite one $\pi_E$ for MP-DDPG. Comparing SGAC in the Figure 3 with MP-DDPG in Figure 2, once the current state $s_k$ is observed, the expert action $a_k^E$ in SGAC that can be computed by MPBE only serves as a guidance label. Then the augmented experience $(s_k, a_k, a_k^E, r_k, s_{k+1})$ is stored into the replay buffer after receiving $r_k$ and $s_{k+1}$ from the environment at each time step. Then these experiences can be used to update the actor networks.

For training the critic-network $Q_\omega^{\pi_\theta}$, Clipped Double Q-Learning method [47] is applied to estimate the action-value $Q^{\pi_\theta}$ as follows,

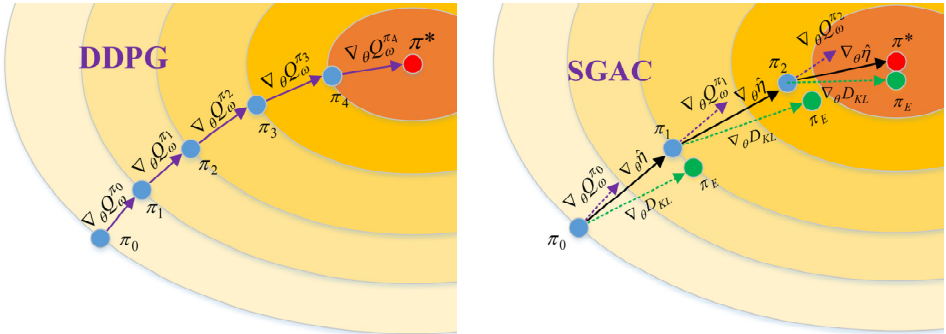

Figure 4. Comparison of policy updating procedure between DDPG and SGAC

$$\omega_j^* = \arg\min_{\omega_j} E_{(s_i,a_i,r_i,s_{i+1})\sim RB}[(y_i - Q_{\omega_j}^{\pi_\theta}(s_i,a_i))^2], \quad j=1,2. \tag{38}$$

And the TD-target $y_i$ in (38) is computed by,

$$y_i = r_i + \gamma \min_{j=1,2} Q_{\bar{\omega}_j}^{\pi_\theta}(s_{i+1}, \mu_{\bar{\theta}}(s_{i+1}) + Z_c), \tag{39}$$

where $Z_c$ is a Gaussian random variable clipped by a positive constant $c$.

All the weights in three target networks, $\bar{\theta}$, $\bar{\omega}_1$ and $\bar{\omega}_2$, are updated using the following manner,

$$\bar{\theta} \leftarrow \varepsilon\theta + (1-\varepsilon)\bar{\theta}, \quad \bar{\omega}_j \leftarrow \varepsilon\omega_j + (1-\varepsilon)\bar{\omega}_j, \quad j=1,2, \tag{40}$$

where the parameter $\varepsilon \ll 1$.

The detailed description of SGAC is given in Algorithm 3. In step 4, the expert action is calculated by MPBE same as MP-DDPG. However, it is worthy to pointing out that the agent action but not the elite one is executed and interact with the environment in step 5. In step 6, the expert action, together with other data, constitutes an augment transition and is stored in RB. In step 9 to 12, all the networks and variables are updated according to the abovementioned procedures. In step 14, the dynamic model used in MPBE is updated through data sampled from RB as described in Section 3.2.

**Remark 4:** Since the critic-network (36) in SGAC is updated using the data $(s_k, a_k, r_k, s_{k+1})$ which are collected by the agent policy itself, rather than using the expert data $(s_k^E, a_k^E, r_k^E, s_{k+1}^E)$ in MP-DDPG, SGAC can estimate $Q^{\pi_\theta}$ according to the standard policy evaluation [11].

**Assumption 2:** For any two consecutive policy parameters $\theta$ and $\theta'$ learned by SGAC and the expert policy $\pi_E$ generated by MPBE in SGAC, the following inequality holds,

$$\max_s D_{KL}(\pi_{\theta'}, \pi_\theta) \leq \max_s D_{KL}(\pi_E, \pi_\theta). \tag{41}$$

---

**Algorithm 3:** Self-Guided Actor-Critic

**Initialize** critic networks $Q_{\omega_1}$, $Q_{\omega_2}$ and actor network $\mu_\theta$, target networks $Q_{\bar{\omega}_1}$, $Q_{\bar{\omega}_2}$ and $\mu_{\bar{\theta}}$ with $\bar{\theta} \leftarrow \theta$, $\bar{\omega}_1 \leftarrow \omega_1$ and $\bar{\omega}_2 \leftarrow \omega_2$, experience replay buffer RB, dynamic model $\hat{f}$, optimization horizon H, the number of control sequences P, dual variable $\lambda$, and task time limit T.

1: **for** each episode **do**
2:    **for** $k = 0, \ldots, T$ **do**
3:       **Observe** current state $s_k$
4:       **Obtain** expert action $a_k^E$ by MPBE and save as a label
5:       **Execute** the agent action $a_k = \mu_\theta(s_k) + Z$, $Z \sim N(0,I)$
6:       **Receive** reward $r_k$ and next state $s_{k+1}$ from the environment
7:       **Store** argument transition $(s_k, a_k, a_k^E, r_k, s_{k+1})$ in RB
8:       **Sample** a random minibatch of N transitions $(s_k, a_k, a_k^E, r_k, s_{k+1})$ from RB
9:       **Update** the critic-networks by (38), (39)
10:      **Update** the actor-network by (34), (35)
11:      **Adjust** the dual variable $\lambda$ by (36), (37)
12:      **Update** the target networks by (40)
13:    **end for**
14:    **Update** dynamic model $\hat{f}$
15: **end for** and **output** trained policy $\pi_\theta$

Since a KL divergence penalty term is imposed during the policy parameter updating for the actor-network, maximum KL divergence between two consecutive policy $\pi_\theta$ and $\pi_{\theta'}$ cannot be arbitrary large. And the inequality (39) is a reasonable condition, namely, the magnitude is bounded by the maximum KL divergence $\max D_{KL}(\pi_E, \pi_\theta)$ between $\pi_\theta$ and $\pi_E$.

**Remark 4:** SGAC takes advantage of imitation learning with a style like DAgger. For each state explored by the agent itself, the expert in SGAC is continuously queried during training to provide a demonstration action. Then the distribution mismatch problem can be addressed since all possible states in the agent trajectories have the guidance of expert demonstration. However, in MP-DDPG, the expert interacts with the environment rather than the agent. And the state that the agent may encounter is not obtainable for the expert. Hence the training data of MP-DDPG is biased from the expert policy.

*4.2. Convergence Analysis of SGAC*

Since most RL algorithms utilize an approximated gradient (surrogate objective) to solve the optimization problem, it is necessary to consider whether the algorithm can find an optimal solution under the true objective function or not. In this section, we theoretically analyze the surrogate objective $\hat{\eta}$ used in (34) - (37) and prove the convergence of SGAC algorithm.

As mentioned in Section 2, an approximated objective $\hat{J}_\theta(\cdot)$ of the true objective $J(\cdot)$ can be used to update the policy parameters when policy gradient method is applied. In [34], the R.H.S. of (15) in Lemma 1 is used to construct a surrogate objective function. And this leads to a conservative learning procedure between consecutive iterations due to the penalty term $\max_s D_{KL}(\pi_{\theta'}, \pi_\theta)$ during optimization. In our paper, however, the surrogate objective $\hat{\eta}$ in (34) - (37) for the proposed SGAC utilizes expert data as guidance. In such circumstance, the learning process of SGAC can be accelerated by imitating the expert policy.

A lemma to evaluate the performance disparity between the agent policy and the expert policy under the true objective $J(\cdot)$ is given when SGAC proposed here is applied to address the RLfD problem (32).

**Lemma 3:** Given an agent policy $\pi_\theta$ and the expert policy $\pi_E$ generated by MPBE mechanism (23), the following inequality holds under Assumption 1,

$$J(\pi_E) - J(\theta) \geq \sigma - \frac{4\gamma\varepsilon_E}{(1-\gamma)^2} \max_s D_{KL}(\pi_E, \pi_\theta), \qquad (42)$$

where $\varepsilon_E = \max_{s,a} |A^{\pi_E}(s,a)|, \forall s \in \mathbf{S}, \forall a \in \mathbf{A}$.

**Proof:** From (11) - (13), we have,

$$\begin{aligned}
&\hat{J}_\theta(\pi_E) - \hat{J}_\theta(\theta) \\
&= \hat{J}_\theta(\pi_E) - J(\theta) \\
&= \int_s \rho_{\pi_\theta}(s) \int_a \pi_E(a|s) A^{\pi_\theta}(s,a) \\
&= \int_s \rho_{\pi_\theta}(s) E_{a \sim \pi_E}[A^{\pi_\theta}(s,a)] \\
&\geq \sigma \cdot \int_s \rho_{\pi_\theta}(s) = \sigma
\end{aligned}$$

where the last inequality comes from Assumption 1. Then from Lemma 1 we have,

$$J(\pi_E) - J(\theta)$$
$$\geq \hat{J}_\theta(\pi_E) - \hat{J}_\theta(\theta) - \frac{4\gamma\varepsilon_E}{(1-\gamma)^2} \max_s D_{KL}(\pi_E, \pi_\theta)$$
$$\geq \sigma - \frac{4\gamma\varepsilon_E}{(1-\gamma)^2} \max_s D_{KL}(\pi_E, \pi_\theta),$$

□

Now we can show policy improvement by SGAC in terms of the true objective $J(\cdot)$ when optimizing the surrogate objective $\hat{J}_\theta(\cdot)$ with the help of the expert data.

**Theorem 2** (Expert Guided Policy Improvement): Under the Assumption 1 and Assumption 2, updating the policy parameter $\theta$ and dual variable $\lambda$ via the surrogate objective $\hat{\eta}(\cdot,\cdot)$ in (34) - (37), SGAC can guarantee the performance $J(\cdot)$ of two consecutive policies is non-decreasing as follows,

$$J(\theta') \geq J(\theta).$$

Proof: Given an expert policy, the following inequalities are derived from Lemma 1 and 3,

$$\left|J(\theta') - \hat{J}_\theta(\theta')\right| \leq \frac{4\gamma\varepsilon_\pi}{(1-\gamma)^2} \max_s D_{KL}(\pi_{\theta'}, \pi_\theta),$$

$$\left|J(\pi_E) - J(\theta)\right| \leq \frac{4\gamma\varepsilon_E}{(1-\gamma)^2} \max_s D_{KL}(\pi_E, \pi_\theta) - \sigma$$

Then we have,

$$\left|J(\theta') - \hat{J}_\theta(\theta')\right|$$
$$\leq \left|J(\theta') - \hat{J}_\theta(\theta')\right| + \left|J(\pi_E) - J(\theta)\right|$$
$$\leq \frac{4\gamma\varepsilon_\pi}{(1-\gamma)^2} \max_s D_{KL}(\pi_{\theta'}, \pi_\theta) + \frac{4\gamma\varepsilon_E}{(1-\gamma)^2} \max_s D_{KL}(\pi_E, \pi_\theta) - \sigma$$
$$\leq \frac{4\gamma(\varepsilon_\pi + \varepsilon_E)}{(1-\gamma)^2} \max_s D_{KL}(\pi_E, \pi_\theta) - \sigma,$$

where the last inequality holds under Assumption 2. And then,

$$J(\theta') \geq \hat{J}_\theta(\theta') - \frac{4\gamma(\varepsilon_\pi + \varepsilon_E)}{(1-\gamma)^2} \max_s D_{KL}(\pi_E, \pi_\theta) + \sigma. \tag{43}$$

Comparing with (15) in Lemma 1, (43) shows the benefits of expert policy imitation learning. Subtracting $J(\theta)$ from both side of (43),

$$J(\theta') - J(\theta) \geq \hat{J}_\theta(\theta') - \hat{J}_\theta(\theta) - \frac{4\gamma(\varepsilon_\pi + \varepsilon_E)}{(1-\gamma)^2} \max_s D_{KL}(\pi_E, \pi_\theta) + \sigma,$$

where $\hat{J}_\theta(\theta)$ is denoted as $Q_\omega^{\pi_\theta}$ in SGAC, and it is an approximation of the true objective $J(\theta)$ via previous data. Applying the updating rule (34) of SGAC can maximize $\hat{J}_\theta(\theta)$ w.r.t. $\theta$, which ensure that $\hat{J}_\theta(\theta') - \hat{J}_\theta(\theta) \geq 0$. And the dual optimization procedure can guarantee the constraint $D_{KL}(\pi_E, \pi_\theta) = 0$ to be satisfied during learning. Thus, $J(\theta') \geq J(\theta)$ holds with a non-negative margin $\sigma$ when the parameter $\theta'$ is learned by SGAC.

□

**Remark 5:** There is not explicit constraint on $D_{KL}(\pi_{\theta'}, \pi_\theta)$ for SGAC, comparing with TRPO or other policy gradient methods [34]. Large parameter updating magnitude can be tolerated by SGAC, rather than taking a penalty by KL divergence for two consecutive policies as [34]. In addition, the constraint of SGAC provides a clear direction for mimicking the expert policy, which also bring of policy improvement margin.

**Corollary 1**: Under the Assumption 1 and Assumption 2, the policy $\pi$ updated in the policy space $\Pi$ by SGAC converges to an policy $\pi_\infty$ satisfying

$$Q^{\pi^*}(s,a) \geq Q^\pi(s,a), \forall s,a,\pi \in \mathbf{S},\mathbf{A},\Pi. \tag{44}$$

Proof: Given a policy $\pi_e$ at episode $e$. Applying standard policy evaluation can get an action-value estimation $Q^{\pi_e}$ of $\pi_e$ [11]. Then the policy monotonic improvement (Theorem 2) can update $\pi_e$ to a new policy $\pi_{e+1}$ such that $Q^{\pi_{e+1}}(s,a) \geq Q^{\pi_e}(s,a)$ holds for all state and actions. In addition, the definition (6) preserve the boundness of $Q^{\pi_e}$. Repeating above two steps can generatesthe policy sequence $\pi_0, \pi_1, ..., \pi_e, ...$

Then from Theorem 1 and 2 we have,

$$Q^{\pi_\infty} \geq Q^{\pi_{\infty-1}} \geq ... \geq Q^{\pi_0},$$

which means $Q^{\pi_\infty}(s,a) \geq Q^\pi(s,a), \forall s,a \in \mathbf{S},\mathbf{A}$ holds for all policy in the sequence. □

## 5. Simulations

To verify the effectiveness of SGAC in dealing with ship berthing control problem, we carry out following simulations for an underactuated ship, and also compare the performance with TD3 [47] and MP-DDPG on the same berthing task.

*5.1. Automatic Ship Berthing Control by SGAC*

In this section, SGAC is used to train an agent to control an underactuated ship to berth in a $10 \times 6$ square meters water area. The ship for simulation is based on the model of a surface vessel with a mass of 17.6 kilograms, a length of $l = 0.6$ meter and the following parameters [14],

$$m_{11} = 19.0, \ m_{22} = 35.2, \ m_{33} = 4.2 \ d_{11} = 4.0, \ d_{22} = 10.0, \ d_{33} = 1.0$$

And all states in **S** are bounded by $s_{\min}$ and $s_{\max}$,

$$s_{\min} = [0\,m/s, 0\,m/s, -5\deg/s, \delta\,m, \delta\,m, 0\deg],$$
$$s_{\max} = [1.0\,m/s, 1.0\,m/s, 5\deg/s,, (10-\delta)\,m, (6-\delta)\,m, 360\deg],$$

where $\delta = 0.5$ meter is the redundancy distance to prevent the ship touch the boundaries of the water area. And all actions in **A** are bounded by $a_{\min}$ and $a_{\max}$, where

$$a_{\min} = [-1\,\text{N}, -1\,\text{Nm}], a_{\max} = [1\,\text{N}, \ 1\,\text{Nm}].$$

In our simulation, the ship is desired to berth at the down-left corner with the final state

$$s_F = [0\,\text{m/s}, 0\,\text{m/s}, 0\,\text{m/s}, 0.65\,\text{m}, 1.0\,\text{m}, 180\deg].$$

All velocities in $s_F$ are zero. The ship is required to stay in a safety berthing zone which is a rectangle cantered in $[X_F, Y_F] = [0.65, 1.0]$ with a length of 1.0 meter and a width of 0.5 meter as shown in Fig. 1.

And three different initial conditions are set for the ship berthing task in order to test the algorithm. In the first case, the ship sets sail at the upright corner of the water area with motionless initial state $s_0$ as,

$$s_0 = [0\,\text{m/s}, 0\,\text{m/s}, 0\,\text{m/s}, 9.0\,\text{m}, 5.0\,\text{m}, 270\deg],$$

where $\psi_0 = 270\deg$ is the initial yaw angle in the earth-fixed frame. In the second case, the ship sets sail at the upright corner of the water area but with a different initial yaw angle,

$$s_0 = [0\,\text{m/s}, 0\,\text{m/s}, 0\,\text{m/s}, 9.0\,\text{m}, 5.0\,\text{m}, 180\,\text{deg}],$$

In the third case, the ship sets sail at the downright corner of the water area with the initial state as,

$$s_0 = [0\,\text{m/s}, 0\,\text{m/s}, 0\,\text{m/s}, 9.0\,\text{m}, 1.0\,\text{m}, 180\,\text{deg}].$$

At last, the reward function is defined as

$$R(s_k, a_k) = \|s_k - s_F\|_2 + \log(\|s_k - s_F\|_2 + 0.001) - 0.1 \cdot \|a_k - a_{k-1}\|_2, \tag{45}$$

for precise positioning. The higher the sum of reward is, the faster and more accurate the ship can approach the desired states.

SGAC is implemented with four feedforward neural networks where the actor-network has one hidden layer with 30 hidden nodes, two critic-networks and dynamic model $\hat{f}$ have two hidden layers with both 100 hidden nodes. All the networks have rectified linear units (ReLU) between any two layers, and a final tanh unit following the output of the actor-network. The noise term $Z$ is sampled from a multivariate Gaussian distribution $N(0, I)$. Other hyper-parameters of SGAC are set as follows,

$$\gamma = 0.9,\ \alpha_{SGAC} = 0.001,\ H = 3,\ M = 10.$$

We conduct 700 episodes of training via SGAC for each case, and test their performances only using the actor-network without the noise term $Z$. Fig.5 depicts the ship berthing trajectories given by SGAC during testing for aforementioned three cases. The ship can successfully berth in the target zone in different initial conditions without hitting the boundary within the task time limits $T=150$ seconds. Without loss of generality, we focus on case one in the rest of the section. Time-dependent trajectories of case one including velocity of surge $u_k$, sway $v_k$ and yaw angle $\phi_k$, and position of $X$, $Y$, and yaw angle $\psi$ are shown in Fig.6-11. It can be clearly found that all velocity trajectories exhibit similar pattern that is increasing and then decreasing to zero as in Fig.6-8. This reflects the same procedure for ship berthing as illustrated in Fig.1, setting up from the initial state, heading for the final state and berthing in the target zone. In addition, the time evolutions of positions and yaw angle present in Fig.6-8 show a smooth berthing process of the ship. Comparing with existing methods for ship berthing, the agent trained by SGAC can accomplish the task without access to the dynamic model information as used in [2][5][48].

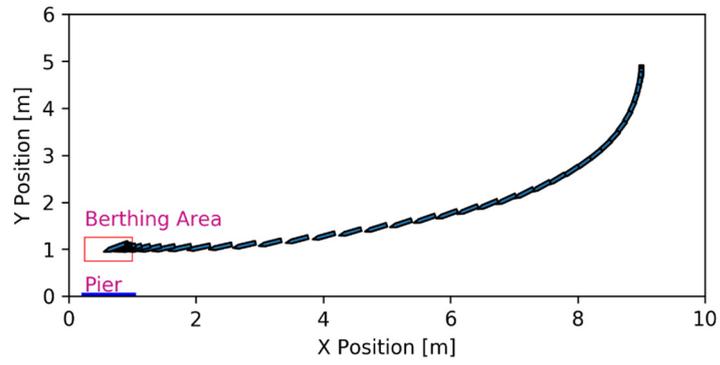

case 1

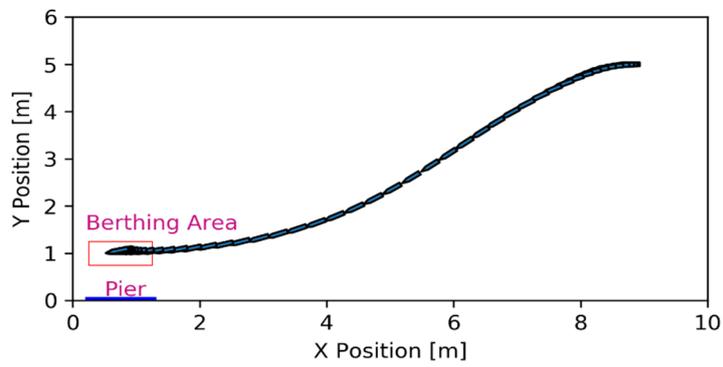

case 2

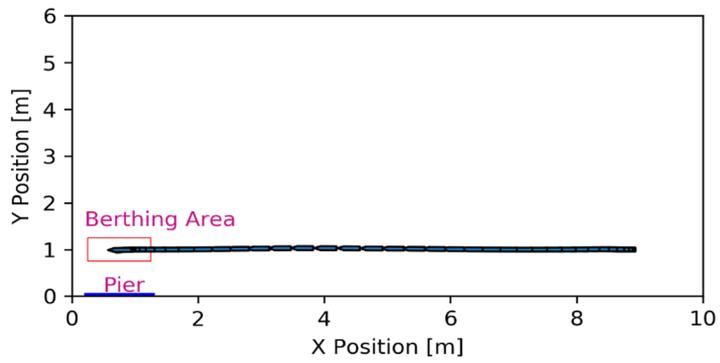

case 3

Figure 5. Test of ship berthing trajectories after training

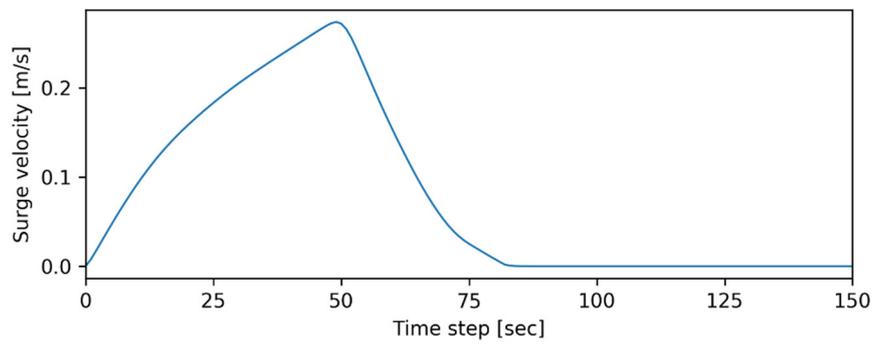

Figure 6. Test of surge velocity trajectory after training

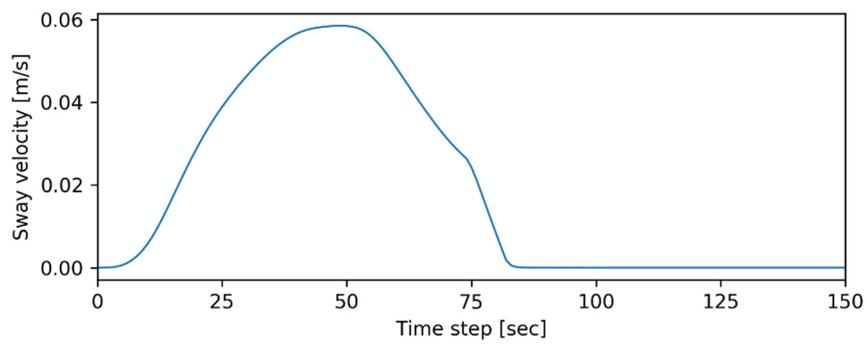

Figure 7. Test of sway velocity trajectory after training

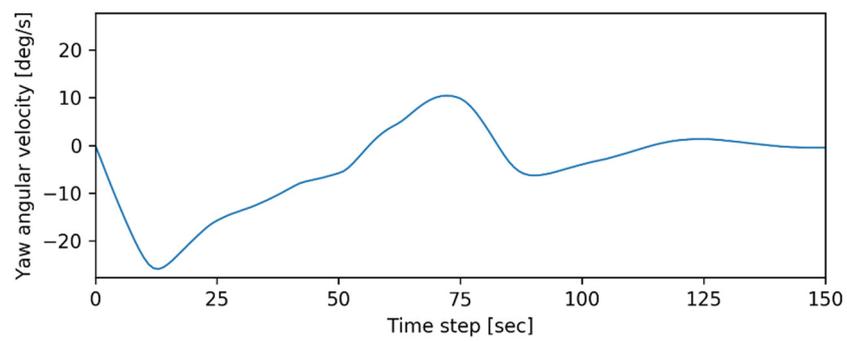

Figure 8. Test of yaw angle velocity trajectory after training

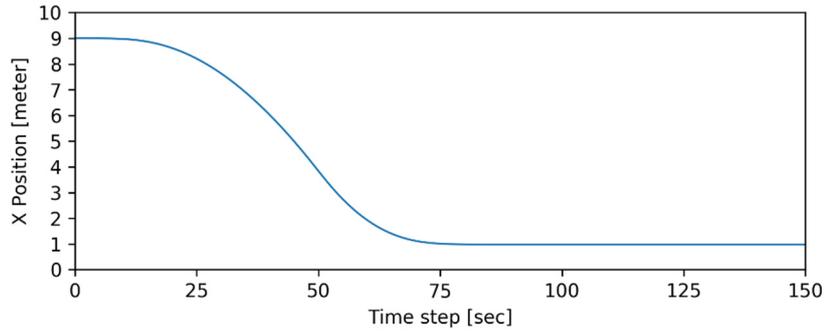

Figure 9. Test of X position of the ship bow after training

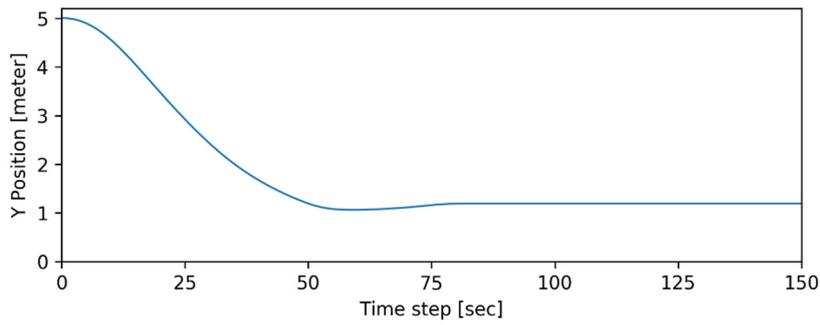

Figure 10. Test of Y position of the ship bow after training

*5.2. Performance Comparison of SGAC with TD3 and MP-DDPG*

To evaluate the effectiveness of SGAC, the same task of case 1 with the same reward function is also simulated by TD3 and MP-DDPG for comparison. Both of TD3 and MP-DDPG have the same implementation setting as SGAC, except $\lambda_{BC} = 0.5$.

Fig.12-15 and Table.1 show the learning processes and testing performances of SGAC, TD3, and two different variants of MP-DDPG (with SM & BC, without SM & BC). As in Fig. 12, the learning process of SGAC has an approximately monotonic improvement, which verifies the conclusion in section 4.

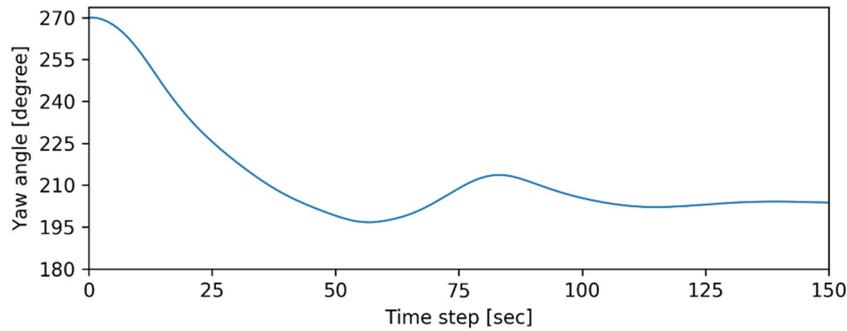

Figure 11. Test of yaw angle after training

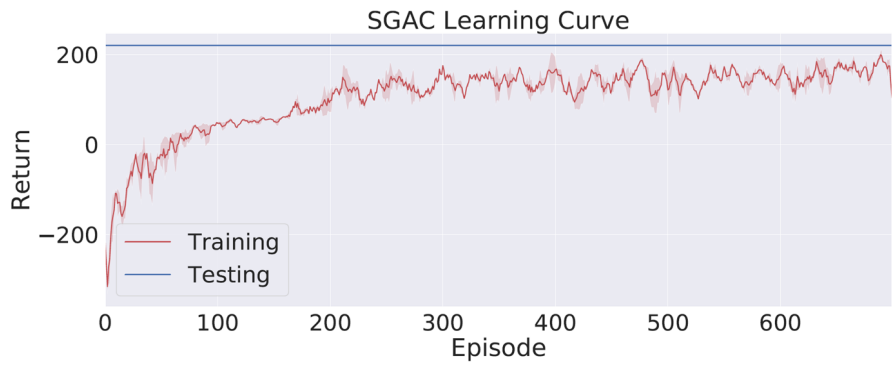

Figure 12. Learning performance of SGAC

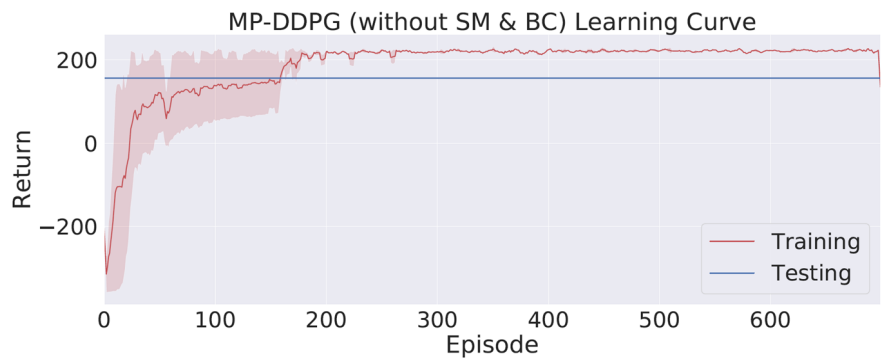

Figure 13. Learning performance of MP-DDPG (without SM & BC)

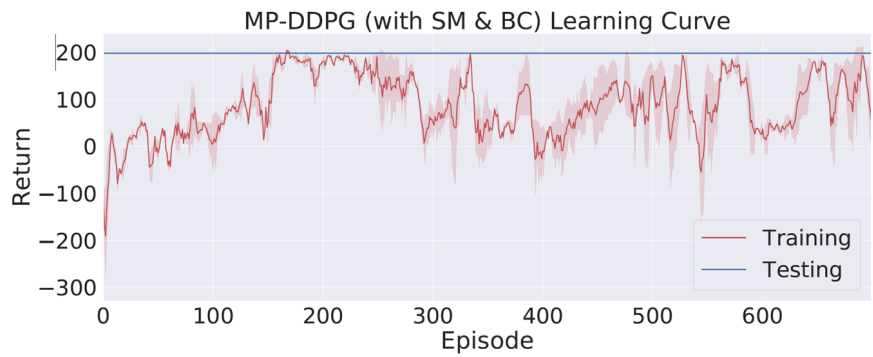

Figure 14. Learning performance of MP-DDPG (with SM & BC)

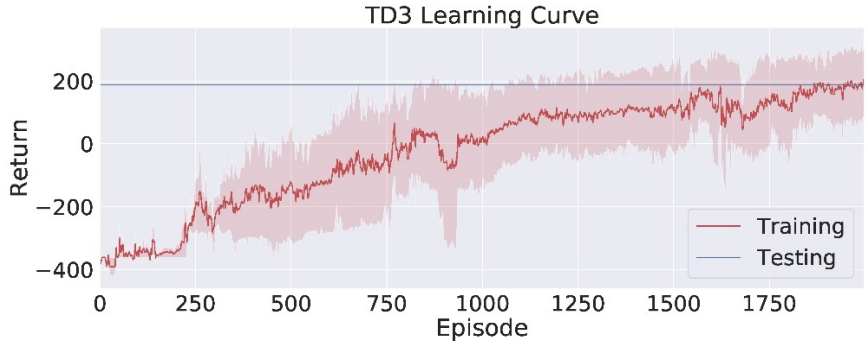

Figure 15. Learning performance of TD3

Table 1. Comparative evaluations

|  | SGAC | MP-DDPG (with SM & BC) | MP-DDPG (without SM & BC) | TD3 |
| --- | --- | --- | --- | --- |
| Maximum training performance | 201.52 | 210.58 | 225.05 | 176.44 |
| Testing performance | 246.19 | 180.15 | 105.17 | 187.15 |

We also conduct 700 episodes of training to two variants of MP-DDPG. Only the expert policy is executed when applying MP-DDPG without SM & BC during training. The learned deterministic policy $\mu_\theta(s_k)$ without the assistance of model predictive mechanism is tested for both variants of MP-DDPG. As shown in Fig.10, MP-DDPG without SM & BC has the fastest training speed comparing with other algorithms, and the expert policy converges after only about 150 episodes. However, the testing performance of MP-DDPG without SM & BC is the worst, and the distribution mismatch problem is serious since that the agent performance during testing is less than half of the expert policy. The result in Fig.14 shows that, although the learning process of MP-DDPG with SM & BC is less stable comparing with the one without SM & BC, the distribution mismatch problem is alleviated to a large extent. Comparing the results shown in Fig.12-15, it is obvious that SGAC has a relatively stable learning process and can well address distribution mismatch problem.

Furthermore, a state-of-art model-free RL method, TD3, is also carried out on the berthing task. Different from the previous work [50], the algorithm runs multiple times here to better exhibit the stability of learning. The agent is trained using TD3 for 2000 episodes, but the learning process is unstable and oscillates even after 1500 episodes (Fig. 15). Comparing with TD3, SGAC can significantly accelerate the convergence speed by leveraging expert demonstrations.

**Conclusion**

In this paper, two novel RLfD algorithms, MP-DDPG and SGAC, which can sufficiently enhance the exploration ability of the agent, are proposed and applied to automatic berthing control problem for underactuated ships with unknown dynamics. In order to generate expert action instead of human or other trained agents, an interactive expert, MPBE, which combines model-free RL and MPC is developed. On the basis of analysing the distribution mismatch problem of MP-DDPG, the expert data are elaborately utilized in SGAC algorithm and the convergence of SGAC are given theoretically. The simulation results verify the effectiveness and advantages of SGAC to execute the ship berthing control task, and also show its ability to overcome the distribution mismatch. In the future, MPBE mechanism should be improved to generate better expert action through more effective optimization techniques rather than random shooting method. The neural network dynamic model used in MPBE can also be replaced by more advance network architecture with the development of Deep learning technology. Furthermore, the practical experiments will be conducted for verification of proposed methods.